\documentclass[twocolumn,aps,prb,epsfig,showpacs,superscriptaddress]{revtex4}

\usepackage{graphicx}%TeX environment specific driver pointer.%

\begin{document}

\title{Full oxide heterostructure combining a high-T$_{C}$ diluted ferromagnet with a high-mobility conductor}

\author{G. Herranz}
\affiliation{Unit\'e Mixte de Physique CNRS / Thales, Route
D\'{e}partementale 128, 91767 Palaiseau, France}
\email{gervasi.herranz@thalesgroup.com}
\author{M. Basletic}
\affiliation{Dep. of Physics, Fac. of Science, HR-10002 Zagreb,
Croatia}
\author{M. Bibes}
\affiliation{Institut d'El\'ectronique Fondamentale, Univ.
Paris-Sud, 91405 Orsay, France}
\author{R. Ranchal}
\affiliation{Depto. F\'isica de Materiales (UCM), Ciudad
Universitaria s/n Madrid 28040, Spain}
\author{A. Hamzic}
\affiliation{Dep. of Physics, Fac. of Science, HR-10002 Zagreb,
Croatia}
\author{E. Tafra}
\affiliation{Dep. of Physics, Fac. of Science, HR-10002 Zagreb,
Croatia}
\author{K. Bouzehouane}
\affiliation{Unit\'e Mixte de Physique CNRS / Thales, Route
D\'{e}partementale 128, 91767 Palaiseau, France}
\author{E. Jacquet}
\affiliation{Unit\'e Mixte de Physique CNRS / Thales, Route
D\'{e}partementale 128, 91767 Palaiseau, France}
\author{J. P. Contour}
\affiliation{Unit\'e Mixte de Physique CNRS / Thales, Route
D\'{e}partementale 128, 91767 Palaiseau, France}
\author{A. Barth\'el\'emy}
\affiliation{Unit\'e Mixte de Physique CNRS / Thales, Route
D\'{e}partementale 128, 91767 Palaiseau, France}
\author{A. Fert}
\affiliation{Unit\'e Mixte de Physique CNRS / Thales, Route
D\'{e}partementale 128, 91767 Palaiseau, France}

\date{\today}

\begin{abstract}

\vspace{0.5cm}

We report on the growth of heterostructures composed of layers of
the high-Curie temperature ferromagnet Co-doped (La,Sr)TiO$_{3}$
(Co-LSTO) with high-mobility SrTiO$_{3}$ (STO) substrates
processed at low oxygen pressure. While perpendicular
spin-dependent transport measurements in STO//Co-LSTO/LAO/Co
tunnel junctions demonstrate the existence of a large spin
polarization in Co-LSTO, planar magnetotransport experiments on
STO//Co-LSTO samples evidence electronic mobilities as high as
$\sim$ 10$^{4}$ cm$^{2}$/Vs at T = 10 K. At high enough applied
fields and low enough temperatures ($\mu$H $\geq$ 6 Teslas, T
$\leq$ 4 K) Shubnikov-de Haas oscillations are also observed. We
present an extensive analysis of these quantum oscillations and
relate them with the electronic properties of STO, for which we
find large scattering rates up to $\approx 10^{-11}$ s. Thus, this
work opens up the possibility to inject a spin-polarized current
from a high-Curie temperature diluted oxide into an isostructural
system with high-mobility and a large spin diffusion length.

\vspace{0.5cm}
\end{abstract}


 \maketitle

\section{Introduction}

The early years of spintonics have focused on the giant
magnetoresistance (GMR) of magnetic metallic
multilayers\cite{baibich88}, and the tunnel magnetoresistance of
magnetic tunnel junctions combining metallic ferromagnets and
insulating barriers \cite{moodera95}. Both have led to interesting
devices such as magnetoresistive read heads, sensors and an
incoming new generation of non volatile magnetic random access
memories (MRAMs).

Further developments in spintronics \cite{zutic2004} now aim at
combining ferromagnets with semiconductors in order to, for
instance, modulate spin-polarized transport by a gate voltage.
This functionality opens exciting perspectives, but is prohibited
in fully metallic structures because of the large carrier density
(n $\approx$ 10$^{22}$ cm$^{-3}$). Unfortunately, at the interface
between a ferromagnetic metal (FM) and a semiconductor (SC)
spin-flip events (inversely proportional to the resistivity $\rho$
times the spin diffusion length $l_{sf})$ lead to an almost
complete loss of the spin polarization inside the metallic
ferromagnet due to the difference in the resistivities between the
two materials\cite{schmidt2000,fert2001}. Such a loss is not
observed in metallic multilayers such as Co/Cu, due to the
similarity of their $\rho\times l_{sf}$ products, and an efficient
spin injection can occur in Cu. It has been suggested that
introducing a spin-dependent resistance at the FM/SC interface can
solve this problem of "conductivity mismatch"
\cite{rashba2000,fert2001}. Yet, one of the best solutions would
be to realize heterostructures combining highly spin-polarized
ferromagnets and non-magnetic materials with similar
resistivities, to enable efficient spin injection (in the ohmic
regime), and low carrier density, to enable gate voltage effects.

Most efforts in this direction have focused on semiconductor
structures using II-VI or III-V diluted magnetic semiconductors
(BeZnMnSe \cite{fiederling99} or (Ga,Mn)As \cite{young2002}) as
spin-polarized injectors. Since the Curie temperature (T$_{\rm
C}$) of (Ga,Mn)As is lower than about 170K \cite{macdonald2005},
such devices are limited to low-temperature operation.

Oxide materials have not been considered so far but should be
interesting candidates since high-T$_{\rm C}$ ferromagnetism has
been reported in diluted magnetic oxide systems (DMOS) like
Co-doped TiO$_{2}$ \cite{matsumoto2001} and ZnO \cite{coey2005}
for instance. In addition, high-T$_{\rm C}$ ferromagnetism was
recently found in La- and Co-doped SrTiO$_{3}$ thin films
\cite{zhao2003,ranchal2005}. In this paper we present results on
structures combining (La, Sr)Ti$_{1-x}$Co$_{x}$O$_{3}$ (Co-LSTO)
and SrTiO$_{3}$ (STO). The archetypical perovskite STO is widely
used as an insulating and diamagnetic substrate for thin film
growth. Remarkably, its transport properties can be drastically
modified by doping with Nb \cite{tufte67}, La \cite{tokura93} or
by creating oxygen vacancies, and a transition towards a metallic
state occurs for carrier densities as low as $\approx$ 10$^{17}$
cm$^{-3}$ (\onlinecite{frederikse64,tufte67}).

Here, we report the properties of ferromagnetic Co-LSTO thin films
grown on STO (001) substrates at low oxygen pressure
(7$\times$10$^{-7}$ $\leq$ P$_{O2}$ $\leq$ 5$\times$10$^{-6}$
mbar). In these conditions, the Co-LSTO films are ferromagnetic
with a large intrinsic spin-polarization and the STO substrates
are conductive. Both materials have low-temperature resistivities
in the m$\Omega \cdot$cm range and in STO, we find mobilities of
up to 10$^{4}$ cm$^{2}$/Vs. Accordingly, Shubnikov-de Haas (SdH)
oscillations are observed and their analysis reveal that carriers
have long scattering rates of $\tau \approx 10^{-11}$ s,
suggesting a spin-diffusion length exceeding 1 $\mu$m. We argue
that this observation, together with the low carrier density in
STO and the large spin-polarization of Co-LSTO, opens the way for
the realization of full oxide spin field effect transistors
(spin-FET).

\section{Experimental}

\begin{table*}[htpb!]
\caption{List of samples analyzed in this
paper}\label{tab:LoS}\vspace{1em}
%\begin{ruledtabular}
\begin{tabular}{ccccc}
  \vspace{0.5em}
  Sample & Substrate & Co\% & $t_{LSTO}(nm)$ & $P_{O2}(mbar)$\\
  \hline
  $\sharp$1 & STO & 0 & 150 & $10^{-6}$\\
  $\sharp$2 & STO & 1.5 & 150 & $8\cdot10^{-7}$\\
  $\sharp$3 & STO & 1.5 & 150 & $10^{-6}$\\
  $\sharp$4 & STO & 1.5 & 20 & $10^{-6}$\\
  $\sharp$5 & LAO & 1.5 & 150 & $7\cdot10^{-7}$\\
  $\sharp$6 & STO & 1.5 & 130 & $2\cdot10^{-4}$\\
\end{tabular}
\vspace{0.5em}
%\end{ruledtabular}
\label{tabsamp}
\end{table*}

\begin{figure}[!h] \vspace{0.3cm}
 \includegraphics[keepaspectratio=true,width=0.9\columnwidth]{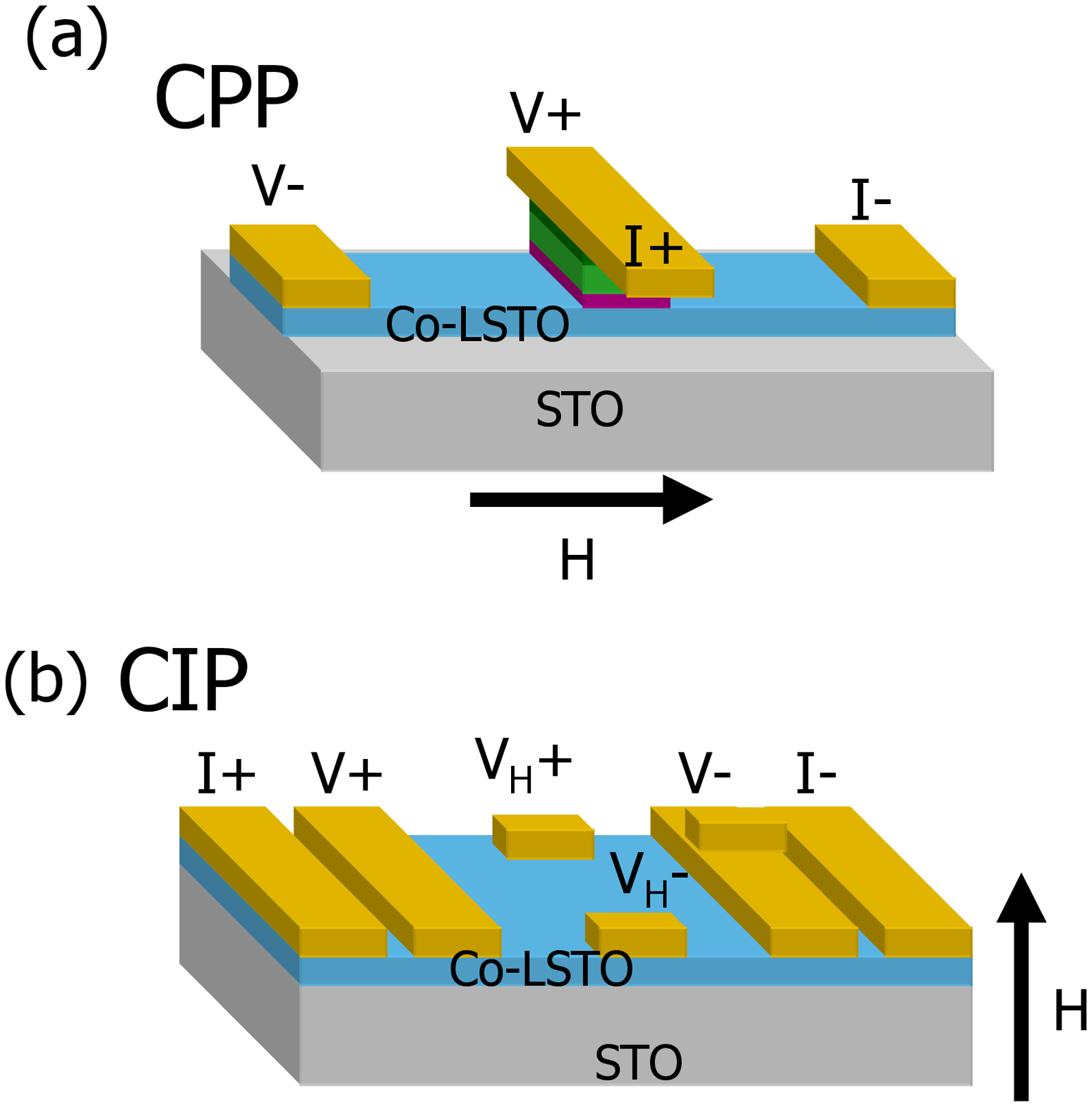}
 \caption{Geometry of the current-in-plane (CIP)
 and current-perpendicular-to-plane (CPP) experiments.
}
 \label{cipcpp}
\end{figure}

Co-LSTO and LSTO epitaxial thin films were grown on 10 mm $\times$
10 mm $\times$ 0.5 mm STO(001) or LAO(001) (LAO : LaAlO$_3$)
substrates by pulsed laser deposition (PLD) \cite{ranchal2005}
from (La,Sr)Ti$_{1-x}$Co$_{x}$O$_{3}$ targets with x = 0.02 and x
= 0 (see Table \ref{tabsamp} for a summary of the sample
description). The La/Sr ratio was about 2:1. The deposition oxygen
pressure (P$_{O2}$) was varied from 7$\times$10$^{-7}$ mbar to
10$^{-4}$ mbar. For current-perpendicular-to-plane (CPP, Fig.
\ref{cipcpp}(a)) transport measurements in Co-LSTO-based magnetic
tunnel junctions (MTJs), a 2.8 nm LAO layer was grown in the same
conditions on the Co-LSTO films. The Co-LSTO/LAO bilayers were
then covered by a Co/CoO/Au stack, processed into tunnel junctions
\cite{bowen2003} and measured in standard 4-wire DC configuration.
The magnetic properties of thin films were measured with an
alternating gradient force magnetometer (AGFM) at room
temperature. Current-in-plane (CIP, Fig. \ref{cipcpp}(b))
magnetotransport measurements were performed on Co-LSTO and LSTO
single films in Hall geometry, with an AC set-up (22 Hz) and in
magnetic fields up to B = 16 teslas and temperatures down to
T=1.5K. Each sample had a pair of current, voltage and Hall
contacts, to which 30 $\mu$m diameter platinum wires were attached
with silver paint. The Hall resistance R$_{xy}$ was measured at
fixed temperatures and in field sweepings from maximum negative to
maximum positive value, in order to eliminate the possible mixing
of magnetoresistive components $(R_{xy} = [R_{xy}(B)-
R_{xy}(-B)]/2)$. The Hall resistance was linear with the magnetic
field for all samples. The mobility was determined as $\mu =
L/w(R_{xy}/R_{xx}B)$, where $L$ is the distance between the
voltage contacts, $w$ is the width of the specimen and $R_{xx}$ is
the CIP-resistance. For the measured samples, $L$ varied between 5
and 6 mm, and $w$ = 2 mm.

\section{Magnetometry and CPP measurements: Ferromagnetism and spin polarization of Co-LSTO}

\begin{figure}[!h] \vspace{0.3cm}
 \includegraphics[keepaspectratio=true,width=0.9\columnwidth]{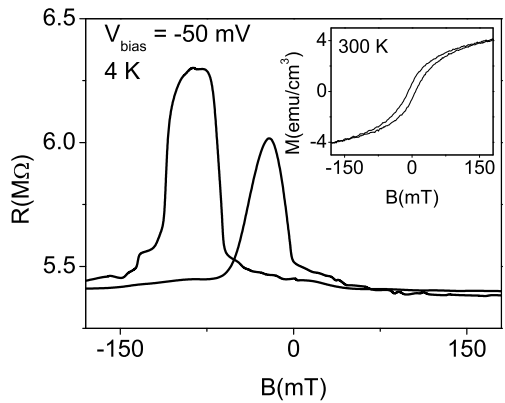}
 \caption{Field dependence of the resistance of a Co-LSTO/LAO/Co magnetic
 tunnel junction (size 64 $\mu$m$^{2}$), at 4K and bias-voltage of 50 mV.
 The inset shows a magnetization hysteresis cycle measured
 at 300K for a Co-LSTO (1.5\%) film grown at $10^{-6}$ mbar.}
 \label{m}
\end{figure}

The inset of Fig. \ref{m} shows that Co-LSTO films are
ferromagnetic at room temperature, with a maximum saturation
magnetization (M$_{S}$) of about 5 emu/cm$^{3}$ obtained for films
grown at the lowest pressures (P$_{O2}$ $\leq$ 5$\times$10$^{-6}$
mbar). We have probed the spin polarization of the Co-LSTO layers
by measuring the magnetotransport properties (in CPP geometry) of
Co-LSTO/LAO/Co MTJs. The existence of a tunnel magnetoresistance
(TMR $\approx$ 20 \%, Fig. \ref{m}) in these tunnelling
experiments demonstrates the spin-polarization of Co-LSTO. The
details of these experiments with, in particular, the analysis of
the peculiar TMR bias dependence and the determination of the spin
polarization (-80\%) are presented elsewhere \cite{herranz2005}.

\section{CIP measurements: resistance and mobility}

\begin{figure}[!h]
\vspace{0.3cm}
 \includegraphics[keepaspectratio=true,width=1\columnwidth]{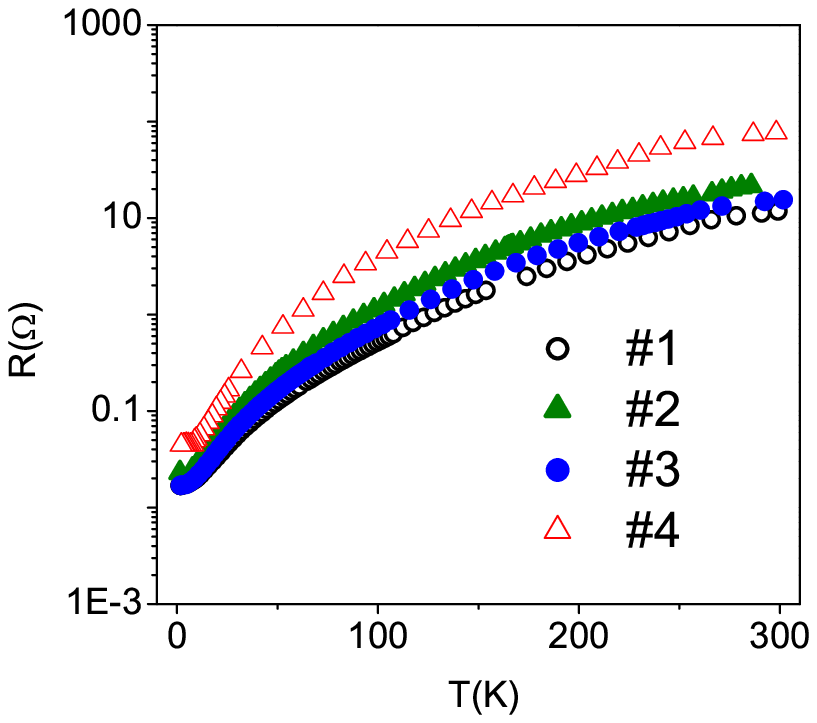}
 \caption{Temperature dependence of the
  raw resistance of samples
 $\sharp1 - \sharp4$.}
 \label{r1}
\end{figure}

\begin{figure}[!h]
\vspace{0.3cm}
 \includegraphics[keepaspectratio=true,width=1\columnwidth]{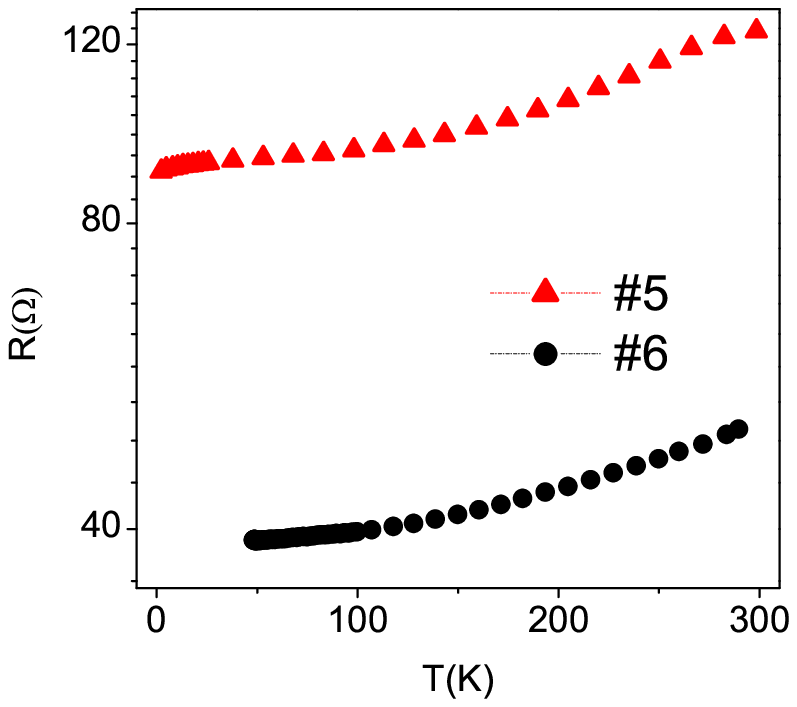}
 \caption{Temperature dependence of the raw resistance of samples
 $\sharp5$ and $\sharp6$.}
 \label{r2}
\end{figure}

\begin{figure}[!h]
\vspace{0.3cm}
 \includegraphics[keepaspectratio=true,width=1.1\columnwidth]{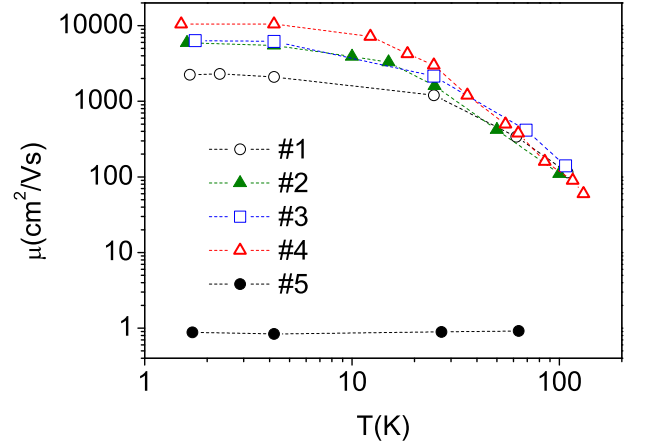}
 \caption{The temperature dependence of the electronic mobilities
  ($\mu_{H}$) for samples grown on
 STO at $P_{O2} < 10^{-6}$ mbar and on LAO ($P_{O2} < 10^{-6}$ mbar).}
 \label{mob}
\end{figure}

We turn now to the CIP measurements of Co-LSTO and LSTO films. The
samples grown on STO at low P$_{O2}$ (samples $\sharp 1 - \sharp
4$) show extremely large $\Re=R_{xx}(300K)/R_{xx}(2K)$ ratios
($\sim$ 700 - 1700, cf. Fig. \ref{r1}) and large electronic
mobilities at low temperature (up to $\mu\approx$ 10$^{4}$
cm$^{2}$/Vs, cf. Fig. \ref{mob}). In contrast, a film grown on LAO
also at low P$_{O2}$ (sample $\sharp 5$, cf. Fig. \ref{r2}) has a
$\Re\approx 1.4$ (i. e. about 3 orders of magnitude lower) and
poor electronic mobility (around 1 cm$^{2}$/Vs at T = 2 - 100 K;
cf. Fig. \ref{mob}). In the same way, a film grown on STO at high
P$_{O2}=10 ^{-4}$ mbar (sample $\sharp 6$, cf. Fig. \ref{r2}) has
also a low ratio $\Re\approx 1.3$ and a mobility $\mu$ = 5.5
cm$^{2}$/Vs.

\begin{table*}[htpb!]
\caption{Transport properties of high-mobility STO
substrates}\vspace{1em}
%\begin{ruledtabular}
\begin{tabular}{cccccc}
  \vspace{0.5em}
  Sample & $\mu_{H}$(1.75 K) & $\Re$ & $n (10^{18} cm^{-3})$ & $m_{H}/m_{e}$
   & $\rho_{2K} (m\Omega cm$)\\
  \hline
  $\sharp$1 & 2250 & 695 & 1.9 & 0.94 & 1.5\\
  $\sharp$2 & 5950 & 935 & 1.1 & 1.2 & 1.0\\
  $\sharp$3 & 6350 & 912 & 1.2 & 1.2 & 0.8\\
  $\sharp$4 & 10500 & 1706 & 1.0 & 1 & 0.6\\
\end{tabular}
\vspace{0.5em}
%\end{ruledtabular}
\label{tabsampSTO}
\end{table*}

\begin{table*}[htpb!]
\caption{Transport properties of Co-LSTO
films}\label{tab:LoS}\vspace{1em}
%\begin{ruledtabular}
\begin{tabular}{ccccc}
  \vspace{0.5em}
  Sample & $\mu_{H}$(1.75 K) & $\Re$ & $n (10^{18} cm^{-3})$ & $\rho_{2K} (m\Omega cm$)\\
  \hline
$\sharp$5 & 0.9 & 1.4 & 9100 & 0.8\\
$\sharp$6 & 5.5 & 1.3 & 1000 & 1.15\\
\end{tabular}
\vspace{0.5em}
%\end{ruledtabular}
\label{tabsampCoLSTO}
\end{table*}

Our resistance and mobility data can therefore be summarized as
follows. The large values of $\Re$ and $\mu$ appear only when
Co-LSTO films are grown on STO substrates at low oxygen pressure
(samples $\sharp1 - \sharp4$). Indeed, this conductive behavior,
together with a high-mobility, is reminiscent of what was reported
for bulk STO single-crystal samples either doped with Nb, treated
in reduced atmospheres \cite{frederikse64,tufte67} or etched by
accelerated Ar ions \cite{reagor2005}. Since the high $\Re$ and
$\mu$ are observed for the films grown in the most reducing
conditions (i.e., promoting the creation of oxygen vacancies), the
high-mobility and low resistance state come most likely from the
STO substrate. On the other hand, LAO or high-P$_{O2}$ STO are
insulators, and the measured transport properties of samples
$\sharp5$ and $\sharp6$ are probably intrinsic of Co-LSTO. Thus,
in our interpretation, the transport properties displayed in Table
\ref{tabsampSTO} (samples $\sharp1 - \sharp4$) refer to the STO
substrate, whereas Table \ref{tabsampCoLSTO} (samples $\sharp5 -
\sharp6$) corresponds to the electronic properties of Co-LSTO.

We note that it is remarkable that the conditions necessary to
grow the DMOS with a large spin-polarization generate this
high-mobility state in the STO. The thickness over which the
high-mobility state is created in STO is discussed later.

\section{CIP measurements: SdH in high-mobility STO covered by Co-LSTO or LSTO}
\label{CIPsection}
\begin{figure}[!h]
\vspace{0.3cm}
 \includegraphics[keepaspectratio=true,width=1.2\columnwidth]{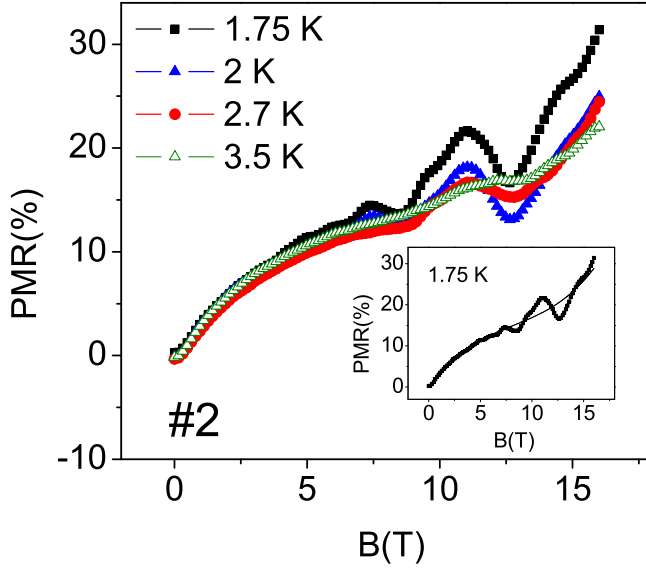}
 \caption{Magnetic field dependence of resistance of sample $\sharp
 2$ at different temperatures, showing SdH oscilations. The inset shows
 the background resistance $R_{bkgnd}$ (solid line) obtained by fitting a
 cubic polynomial.}
 \label{oscRdH}
\end{figure}

 Magnetic field and temperature dependence of the CIP resistance
$R_{xx}$ for sample $\sharp 2$ is shown in Fig. \ref{oscRdH}. The
magnetic field $H$ was applied perpendicularly to the sample
plane. A large positive perpendicular magnetoresistance (PMR =
${(R_{xx}(H)-R_{xx}(0))}/{R_{xx}(0)})$ is observed, showing
oscillations (at T $\leq$ 4 K and $\mu$H $\geq$ 6 T) that are due
to the Shubnikov-de Haas (SdH) effect. We emphasize that we have
observed SdH oscillations in all other high-mobility samples.
Oscillations with the same period are also observed when the field
is applied parallel to the sample plane in the longitudinal
configuration (not shown). Such observation evidences that the
high-mobility region is 3D-like and is not an interface effect.
Quantum transport effects like SdH oscillations are usually
observed in systems with high mobilities. This is the case of
semi-metals like Bi \cite{lerner63} or semiconductors like GaAs
\cite{vul76}, but SdH oscillations have been very seldom observed
in oxide materials \cite{frederikse67,mackenzie98,cao2003}.

\begin{figure}[!h]
\vspace{0.3cm}
 \includegraphics[keepaspectratio=true,width=1\columnwidth]{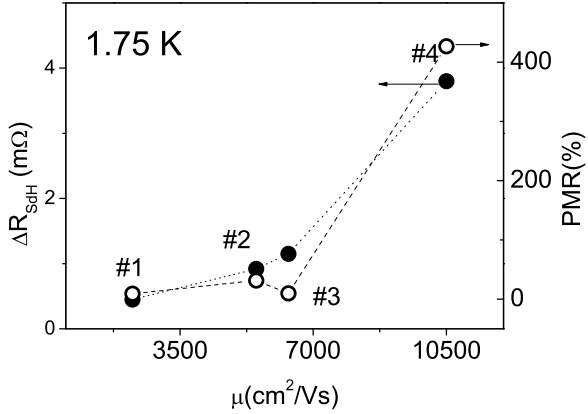}
 \caption{The mobility dependence of the SdH-oscillation
  amplitude $\Delta R_{SdH}$
 and the PMR (for
 B = 16 T)  for samples $\sharp 1 - \sharp 4$.}
 \label{PMR&mu}
\end{figure}

In order to isolate the SdH oscillations, first a background
positive magnetoresistance $R_{bkgnd}$ was determined by fitting
the experimental data $R_{xx}$ to a low-order polynomial (cf.
inset of Fig. \ref{oscRdH}). The  oscillating part $\Delta
R_{SdH}$ of the magnetoresistance was then obtained as $\Delta
R_{SdH} = R_{xx}-R_{bkgnd}$. Fig. \ref{PMR&mu} shows that $\Delta
R_{SdH}$ scales with mobility $\mu$  in the same way as does the
perpendicular magnetoresistance PMR. The origin of the large
values of the PMR is presumably related to the perturbation of the
electronic orbitals by the applied field and, thus, they are more
significant when the mean free path, i. e. the mobility, is
larger. Our results indicate that both SdH amplitudes and PMR
originate from similar physical mechanisms.

\begin{figure}[!h]
\vspace{0.3cm}
 \includegraphics[keepaspectratio=true,width=1\columnwidth]{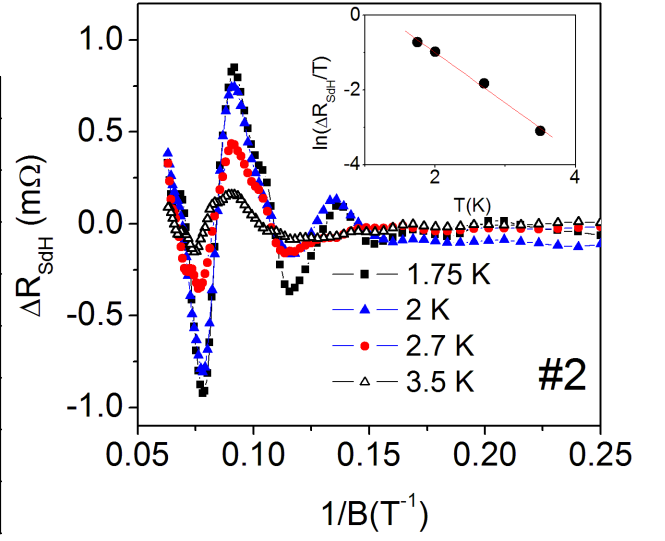}
 \caption{SdH-oscillation
 amplitudes $\Delta R_{SdH}$ vs.
   1/B for sample $\sharp 2$ at different temperatures.
   The inset shows the extraction
 of the effective mass from the
 ln$\Delta R_{SdH}/T$ vs. T plot.}
 \label{oscSdH}
\end{figure}

Fig. \ref{oscSdH} presents the oscillations of $\Delta R_{SdH}$ as
a function of 1/$B$ and temperature. We analyzed these SdH
oscillations in our samples using the conventional
Lifshitz-Kosevich expression \cite{shoenberg84}. The amplitude of
the oscillation is given as:

\begin{equation}
    \Delta R_{SdH} \propto B^{1/2}R_{T}R_{D}R_{S}sin\bigg[2\pi\bigg(\frac{F}{B}-\frac{1}{2}\bigg)\pm \frac{\pi}{4}\bigg]
    \label{LK}
\end{equation}

\noindent where $\Delta R_{SdH}$ is the amplitude of the
SdH-oscillating part of the resistance, and $F = \hbar A/2 \pi e$
($A$ is the extremal orbit area in k-space perpendicular to the
applied field) is expressed in teslas. The oscillation amplitude
is damped due to the smearing of the Fermi surface orbit by
temperature ($R_{T}$), scattering ($R_{D}$) and spin-splitting
effects ($R_S$) :

\begin{equation}
    R_{T} = \frac{\alpha (m_{H}/m_{e}) T/B}{sinh(\alpha (m_{H}/m_{e}) T/B)}
\end{equation}

\noindent with $\alpha=(2\pi^{2}k_{B}m_{e})/{e\hbar} = 14.69$
(TK$^{-1}$);

\begin{equation}
    R_{D} = exp \bigg(\frac{-\alpha (m_{H}/m_{e})T_{D}}{B}\bigg )
\end{equation}

\noindent where $m_{H}$ is the effective mass, $T_{D}$=$\hbar/(2\pi
k_B \tau)$ is the Dingle temperature and $\tau$ is the scattering
rate;

\begin{equation}
 R_S=cos[(\pi/2)g_{eff}(m_H(\theta)/m_{e})]
\end{equation}

\noindent where $g_{eff}$ is the effective Land\'e factor.

\begin{figure}[!h]
\vspace{0.3cm}
 \includegraphics[keepaspectratio=true,width=0.9\columnwidth]{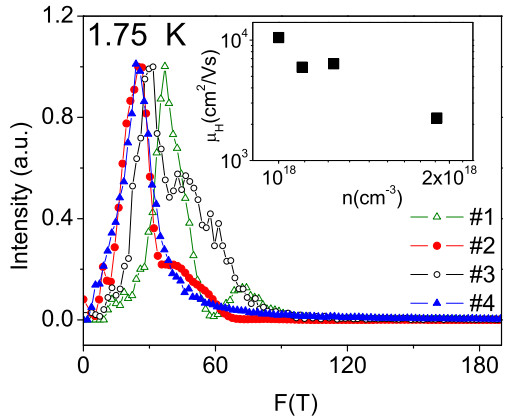}
 \caption{Power Spectrum Intensity of SdH oscillations at T = 1.75 K of samples $\sharp 1 - \sharp 4$.
 The inset shows the evolution of the Hall mobility with the carrier density
 (extracted from the frequency of the SdH oscillations).}
 \label{fft}
\end{figure}

A Fast Fourier Transform (FFT) method has been used to obtain the
power spectra of the measured resistance values. This analysis
indicates a primary frequency of $F \approx 24 - 35$ T, depending
on sample (Fig. \ref{fft}). The data also suggest the presence of
higher-frequency components; however, their amplitude is much
lower, and therefore  they will be neglected in the following
discussion. From the frequency $F$ we can infer the
cross-sectional area $A$ of the closed electronic orbits
contributing to the SdH oscillations. Experimental data support a
conduction band of STO consisting of ellipsoids of revolution
having the long axis along the $<$100$>$ crystalline axes, and
minima at the $\Gamma$-points \cite{kahn64, frederikse64,
mattheiss72}. Taking into account this k-space geometry and the
value of k$_{F}$, we can estimate the carrier density as $n
\approx 1.0\times10^{18} - 1.9\times10^{18}$ cm$^{-3}$ (see Table
\ref{tabsampSTO} and section \ref{annex}). We observe that the
mobility increases with decreasing $n$ (cf. inset of Fig.
\ref{fft}), in agreement with previous reported works
\cite{tufte67}. This may be explained by assuming that oxygen
vacancies act as carrier donors, being simultaneously scattering
centers.

In comparison, the carrier density of sample $\sharp$5 (Co-LSTO on
LAO) deduced from Hall experiments is $n \approx 9 \times 10^{21}$
cm$^{-3}$. We recall that in this case the Hall experiments allow
the extraction of the electronic properties of the film, since the
substrate is an insulator. The measured carrier density is
consistent with about 0.6 carrier per formula unit, which is in a
good agreement with what is expected from the film composition,
assuming the presence of some oxygen vacancies.

For the high-mobility samples ($\sharp 1 - \sharp 4$), the
effective mass ($m_{H}$) of the carriers in STO was estimated by
analyzing the temperature dependence of the SdH oscillations
amplitude and plotting ln $(\Delta R_{SdH}/T)$ vs. temperature.
This expression is accurate within an error $<$ 2\% and can be
deduced from Eq. (\ref{LK}), assuming that T/B $\geq$
0.14/($m_{H}/m_{e})$ ($\rm KT^{-1}$). We have checked this
condition after the extraction of $m_{H}$. Our analysis yields the
values of $m_{H}$ between $0.94 - 1.2 m_{e}$ (cf. Table
\ref{tabsampSTO}). The Dingle temperature $T_{D}$ was obtained
from the slope of the plot of ln $(\Delta
R_{SdH}B^{1/2}$sinh$(\alpha (m_{H}/m_{e})T/B))$ versus 1/B at T =
1.75 K. Due to the limited number of oscillations in the explored
range of
 field values, this analysis turned out to be quite difficult.
 Nevertheless, values of $T_{D} = 0.3 - 1$ K were obtained, corresponding
to scattering rates $\tau_{D} = 4-12\times10^{-12}$ s. The values
of $\tau_{D}$ correspond to the scattering rates of the closed
electronic orbits around the extremal cross sections of the Fermi
surface. Due to the cubic symmetry of STO, we can estimate the
scattering rate $\tau = \frac{\mu m_{H}}{e}$ at low temperature,
assuming that the effective mass is isotropic and equal to
m$_{H}$. Using the values of $\mu$ and m$_{H}$ (table
\ref{tabsampSTO}) we get $\tau = 1.5-6\times10^{-12}$, in good
agreement with the values of $\tau_{D}$.

Knowing the carrier densities and mobilities, we can calculate the
resistivity through $\rho_{xx} = 1/(ne\mu)$. We find increasing
values in the range $\rho_{xx} = 0.6 - 1.5$ m$\Omega$ cm, for
decreasing mobilities within the range 2250 - 10500 cm$^{2}$/Vs
(see Table \ref{tabsampSTO}). These values of $\rho_{xx}$ are
consistent with those from the literature, for similar mobility
values \cite{tufte67}.

The values of $\rho_{xx}$ and $n$ allow us to estimate the order
of magnitude of the thickness $t\rm{_{STO}}$ of the high mobility
region inside the STO. From  $\rho_{xx} = (w
t\rm{_{STO}})R_{xx}/L$, with $L \approx 5-6$ mm and $w \approx 2$
mm, we infer $t\rm{_{STO}} \approx$ 300 - 900 $\rm \mu$m. We can
also estimate $t\rm{_{STO}}$ from the measured Hall constant
$R_{H} = 1/ne = R_{xy} t\rm{_{STO}}/B$, and assuming the values of
the carrier densities extracted from the SdH analysis (Table
\ref{tabsampSTO}) we find again $t\rm{_{STO}} \approx$ 300 - 900
$\rm \mu$m. In other words, the high mobility system is not an
interfacial interface effect, but it is extended to a sizable part
of the STO substrate. This is also supported by the observation of
SdH oscillations when the magnetic field is applied in the film
plane in the longitudinal configuration (see section
\ref{CIPsection}). An additional support for this conclusion comes
from the estimation of  the spread of the oxygen vacancy diffusion
in STO, by using  the diffusion constants determined
experimentally from optical experiments \cite{denk97}. Even in the
less favorable case (considering trapping effects due to
impurities inside the STO, and at pressures $P_{O2} \approx
10^{-2}$ mbar,  i. e.  around four orders of magnitude higher than
our growth conditions), the determined diffusion constants were $D
\approx 10^{-4}$ cm$^{2}$/s at T $\approx 700^{o}$C; with this
value, the diffusion of O-vacancies during the deposition time
$t_{dep} \approx 10^{2} - 10^{3}$ s can be estimated to be $l_{O
vac} \approx (D\times t_{dep})^{1/2} \approx 1$ mm. Thus, the
oxygen vacancies can diffuse across the STO substrate, which is
consistent with the $t\rm{_{STO}} \approx$ 300 - 900 $\rm \mu$m we
found from transport data.

\section{Perspectives}

The resistivity of the Co-LSTO films can be calculated from the
resistance of the LAO//Co-LSTO and high-P$_{O2}$ STO//Co-LSTO
samples and the Co-LSTO film thickness. This yields 0.9
m$\Omega\cdot$cm and 1.15 m$\Omega\cdot$cm at low temperature,
respectively, see Table \ref{tabsampCoLSTO}. Thus, the
low-temperature resistivities of the Co-LSTO films and the
oxygen-deficient STO substrates are in the m$\Omega \cdot$cm
range. Due to the high mobility of the latter, it is tempting to
consider the possibility of spin-injection from Co-LSTO into STO,
as in the Co/Cu case. More precisely, in this case the
spin-polarization of the injected current is given by
\cite{fert2001}

\begin{equation}
P = \beta/(1+r_{N}/r_{F})= \beta/[1+(\rho_{N}l_{sf}^{N} /\rho_{F}l_{sf}^{F})]
\end{equation}

\noindent with $\beta$ being related to the spin-up resistivity
$\rho_\uparrow$ and spin-down resistivity $\rho_\downarrow$
channels in the ferromagnet by $\beta =(\rho_\uparrow -
\rho_\downarrow)/(\rho_\uparrow + \rho_\downarrow)$. $\rho_{F}$,
$\rho_{N}$ and $l_{sf}^{F}$, $l_{sf}^{N}$ are the resistivities
and the spin diffusion length in the ferromagnet and the
non-magnetic material, respectively. In the present case, we know
$\rho_{N} / \rho_{F} \approx 1$ and $\beta$ should be close to -1
as the spin-polarization of Co-LSTO is -80\% \cite{herranz2005}.
In conventional metals, $l_{sf}$ is usually much longer than the
mean free path $\lambda$. Here $\lambda_{STO} \geq 100$ nm and
$\lambda_{Co-LSTO} \approx 1$ nm but the spin diffusion lengths
are unknown for virtually all oxide materials.

Since oxygen-deficient STO thin films with mobilities and
resistivities comparable to those reported for single crystals
have already been fabricated \cite{leitner98,pallecchi2001},
Co-LSTO/STO superlattices could exhibit a significant CPP GMR and
systematic GMR studies should allow measuring $l_{sf}$ in both
materials.

It must be pointed out that the efficient spin-injection from
Co-LSTO into STO may be limited to the lowest temperatures due to
the steep increase of resistivity of STO (cf. Fig. \ref{r1}).
However, if the $r_{N}/r_{F}$ ratio turns out to be too large to
allow an efficient spin-injection, the latter could be increased
by inserting for instance, an ultrathin tunnel barrier of LAO
\cite{fert2001}. Our TMR measurements indeed demonstrate that in
the low oxygen pressure conditions required to obtain
ferromagnetic Co-LSTO and high-mobility STO, LAO retains its
insulating properties.

Finally, we would like to mention that several groups have
fabricated FETs using STO as the channel material, with gains
reaching 100 at room temperature \cite{ueno2003}. Remarkably, the
field effect proves efficient for a wide range of carrier density,
from the undoped \cite{ueno2003} to the metallic state ($n \approx
10^{18}$ cm$^{-3}$) \cite{pallecchi2001} and even in the
low-temperature superconducting phase \cite{takahashi2004}. If the
field effect can modify the spin-diffusion length (in addition to
the carrier density), or enables the tuning of spin-flip
mechanisms (in analogy to the principle of the Datta and Das
transistor \cite{datta90}), full oxide spin-FET using STO channels
and Co-LSTO as spin injector and detector could be built in the
near future.

\section{Conclusion}

In conclusion, we have demonstrated the feasibility of growing a
full oxide structure combining a highly spin polarized diluted
magnetic system Co-LSTO and  high-mobility (up to $\approx 10^{4}$
cm$^{2}$/Vs) STO, by using low oxygen pressure growth conditions.
At low temperature\textbf{s} and high magnetic fields, we have
observed Shubnikov-de Haas oscillations in the resistance of the
STO. Careful analysis of this behavior for several samples has
allowed us to extract valuable information on the electronic
properties of the high-mobility STO. Besides the very large
mobility, we have found that the metallic state of STO occurs for
carriers densities as low as $\sim10^{18}$ cm$^{-3}$, with large
scattering rates up to $\approx 10^{-11}$ s. The combined
properties of the two materials -i. e. the large spin-polarization
of Co-LSTO and the long mean free path of STO - makes them
interesting to realize spin-injection devices like spin FET, the
low carrier density of STO enabling external control by gate
voltage. Several studies have recently demonstrated the potential
of Ti perovskites for oxide electronics
\cite{ohtomo2002,ohtomo2004,okamoto2004}. We are convinced that
our findings extend this potential to oxide spintronics.

\section{Annex: estimation of the carrier density from SdH oscillations}
\label{annex} The Fermi surface of STO is composed of three
ellipsoids of revolution centered at $\Gamma$-points with the long
axis $\rm k_{F,max}$ along $<$100$>$ crystallographic axes and the
short $\rm k_{F,min}$ axes transverse to them \cite{kahn64,
frederikse64, mattheiss72}. The ratio between the longitudinal
(m$_{l}$) and transverse (m$_{t}$) masses have been determined
experimentally as m$_{l}$/m$_{t}$ $\approx$ 4 \cite{kahn64,
frederikse64}. Recalling that $\partial^2 E(\textbf{k})/
\partial\textbf{k}^2 \propto 1/m$, we can infer that $\rm
k_{F,max}$/$\rm k_{F,min} \approx 4$.

Let us assume that a field $\rm H_{z}$ is applied along the [001]
direction. Then the extremal electronic closed orbits
corresponding to the 2 spheroids directed along x- and y-axis are
ellipsoidal with cross sections equal to $\pi\cdot\rm
k_{F,min}\cdot\rm k_{F,max}$, whereas the cross sections of the
ellipsoid along the z-direction is circular with an area
$\pi\cdot\rm k_{F,min}^2$. In addition, since $\rm k_{F,max}$/$\rm
k_{F,min} \approx 4$ there are about 4 more electrons around the
ellipsoidal closed orbits than those orbiting the circular orbits.
Taking into account these facts, we assume that the main frequency
peak of the $\Delta R_{SdH}$ oscillations comes from electrons
orbiting the ellipsoidal cross sections, and, thus, A =
$\pi\cdot\rm k_{F,min}\cdot\rm k_{F,max}$. As the total volume in
the k-space occupied by the 3 ellipsoids is V$_{\textbf{k}}$ =
$3\times\frac{4\pi}{3}$$\rm k_{F,min}^2$$\rm k_{F,max}$, we can
estimate the carrier concentration as $n = 3\times\frac{\rm
k_{F,min}^2 \rm k_{F,max}}{3\pi^{2}}$.

\acknowledgements{We are grateful to P. Berthet, C. Pascanut and
N. Dragoe for providing the Co-LSTO target.  G. Herranz
acknowledges financial support from Minist\'ere de l'Education
Nationale, de l'Enseignement Sup\'erieur et de la Recherche
(France) and R. Ranchal thanks Universidad Complutense de Madrid
and the Spanish Project MAT 2001-3554-CO2 for partially supporting
her stay at UMR137, CNRS-Thales at Orsay, France. The financial
support from the PAI-France-Croatia COGITO program n$^{o}$
82/240083 is also acknowledged. }

\vspace{0.5em}

\bibliographystyle{prsty}

\end{document}